# Indicators of the Interdisciplinarity of Journals:

# Diversity, Centrality, and Citations

Loet Leydesdorff [i] & Ismael Rafols [ii]

**Abstract**

A citation-based indicator for interdisciplinarity has been missing hitherto among the set of available journal indicators. In this study, we investigate network indicators (betweenness centrality), journal indicators (Shannon entropy, the Gini coefficient), and more recently proposed Rao-Stirling measures for "interdisciplinarity." The latter index combines the statistics of both citation distributions of journals (vector-based) and distances in citation networks among journals (matrix-based). The effects of various normalizations are specified and measured using the matrix of 8,207 journals contained in the *Journal Citation Reports* of the (Social) Science Citation Index 2008. Betweenness centrality in symmetrical (1-mode) *cosine*-normalized networks provides an indicator outperforming betweenness in the asymmetrical (2-mode) citation network. Among the vector-based indicators, Shannon entropy performs better than the Gini coefficient, but is sensitive to size. *Science* and *Nature*, for example, are indicated at the top of the list. The new diversity measure provides reasonable results when (1 – *cosine*) is assumed as a measure for the distance, but results using Euclidean distances were difficult to interpret.

---

[i] Amsterdam School of Communication Research (ASCoR), University of Amsterdam, Kloveniersburgwal 48, 1012 CX Amsterdam, The Netherlands; loet@leydesdorff.net ; http://www.leydesdorff.net.
[ii] SPRU – Science and Technology Policy Research, University of Sussex, Brighton, BN1 9QE, England; i.rafols@sussex.ac.uk.

**Introduction**

Among the various journal indicators based on citations, such as impact factors, the immediacy index, cited half-life, etc., a specific indicator of interdisciplinarity has hitherto been lacking (Kajikawa *et al*., 2009; Porter *et al.,* 2006 and 2007; Wagner *et al.*, 2009; Zitt, 2005). Journals obviously vary in terms of the range of disciplines they cover. In their early studies, Francis Narin and his colleagues already noted that *Science* and *Nature* did not fit into their hierarchical clustering scheme (Narin *et al*., 1972; Carpenter & Narin, 1973). At lower levels of the journal hierarchy, however, journals may also fulfill inter- or transdisciplinary functions. For example, *Limnology & Oceanology* has been considered by Leydesdorff (1986) as a journal that overarched the domains of fresh-water and marine biology, and therefore had a higher impact factor.

Given the matrix of aggregated journal-journal citations as derived from the *Journal Citation Reports* (*JCR*) of the (*Social*) *Science Citation Index*, a clustering algorithm usually aims to partition the database in terms of similarities in the distributions— in this case, distributions of citations. Some journals reach across boundaries because they relate different subdisciplines into a single (disciplinary) framework. For example, the *Journal of the American Chemical Society* (*JACS*) accepts contributions to organic, inorganic, physical, analytical chemistry, etc. Other journals combine intellectual contributions based on methods or instruments used in different disciplines. Thus, a network is woven in which relatively distinguishable clusters can be considered as representations of specialties, while hierarchies across specialties can be shaped in different directions. Note that these hierarchies may not only operate on individual journals, but on sets—representing specialties—in which one can construct a center (or "centroid"). A generalized measure of interdisciplinarity at the journal level therefore is far from obvious.

Furthermore, interdisciplinarity may be a transient phenomenon. As a new specialty emerges, it may draw heavily on its mother disciplines/specialties, but as it matures a set of potentially new journals can be expected to cite one another increasingly, and thus to develop a type of closure that is typical of "disciplinarity" (Van den Besselaar & Leydesdorff, 1996). Interdisciplinarity, however, may mean something different at the top of the journal hierarchy (as in the case of *Science* and *Nature*) than at the bottom, where one has to draw on different bodies of knowledge for the sake of the application (e.g., in engineering). Similarly, in the clinic one may be more inclined to integrate knowledge from different specialties at the bedside than a laboratory where the focus is on specialization and refinement.

In the bibliometric tradition, therefore, "interdisciplinarity" has remained a difficult issue despite its high policy-relevance (Laudel & Origgi, 2006; Wagner *et al.*, 2009). More recent additions to the database (that is, the *JCR*) such as eigenfactor and article influence scores (Bergstrom, 2007; Rosvall & Bergstrom, 2008) focus heavily on weighing journals in terms of their hierarchical status (cf. Bollen *et al.*, 2006). Most of these new measures rank journals like the PageRank in Google, using information from the entire network of journal-journal citation relations (Page *et al.*, 1998; cf. Pinski & Narin, 1976)

However, none of these new indicators hitherto focused specifically on interdisciplinarity with perhaps the exception of studies of betweenness centrality (Bollen *et al*., 2009a and b; Leydesdorff, 2007 and 2009).

Among the network indicators, betweenness centrality seems an obvious candidate for the measurement of interdisciplinarity (Freeman, 1977, 1978/9). One of us experimented with betweenness centrality as an indicator of interdisciplinarity in aggregated journal-journal citation networks (Leydesdorff, 2007). Large journals (e.g., *Nature* and *Science*), however, may show high betweenness centrality because of their degree centrality: a star in a network can be expected to contribute to a relatively large percentage of the possible pathways among other nodes (Bollen *et al*., 2009a, at p. 6). This can be corrected by normalizing the vectors (rows or columns), for example, by using the *cosine* measure (Salton & McGill, 1983; Ahlgren *et al*., 2003; Bonacich, *personal communication*, 22 May 2006). In *local* environments, betweenness centrality could then be shown to perform outstandingly as an indicator of "interdisciplinarity." However, the usefulness of this indicator at the level of the set has not yet been tested. Using rotated factor analysis, Bollen *et al.* (2009b, at pp. 4 ff.) found betweenness centrality positioned near the origin of a two-factor solution; this suggests that betweenness centrality might form a separate (third) dimension in their array of 39 possible journal indicators.

The occasion for returning to the research question of a journal indicator for "interdisciplinarity" was provided by the new interest in "interdisciplinarity" in bibliometrics (Laudel & Origgi, 2006; Wagner *et al*., 2009) and the availability of another potential measure: diversity as defined by Stirling (2007; cf. Rao, 1982). Would it perhaps be possible to benchmark the various possible indicators of "interdisciplinarity" against each other? Using this new measure, Porter & Rafols (2009) and Rafols & Meyer (2010), for example, suggested that this new measure would be useful to indicate interdisciplinarity at the article level. Might it also provide us with a useful indicator at the journal level?

Stirling (2007, at p. 712) proposed to integrate the (in)equality in a vector with the network structure using the following formula for diversity $D$:

$$D = \sum_{ij(i \neq j)} p_i \cdot p_j \cdot d_{ij} \qquad (1)$$

This measure is also known in the literature as "quadratic entropy" (e.g., Izsáki & Papp, 1995) because unlike traditional measures of diversity such as the Shannon entropy and the Gini index, the probability distributions ($p_i$ and $p_j$) of the units of analysis (in our case, the citation distributions of the individual journals) are multiplied by the distance in the (citation) network among them ($d_{ij}$). The latter factor represents the journal ecology: a journal which is interdisciplinary within a specific domain (e.g., chemistry) may not reach beyond the confines of this domain. Other journals can be highly specialist, but yet combine citations from or to different domains. Thus, diversity has these two aspects which can also be studied separately as the properties of the citation distribution of each journal (e.g., the Gini index) versus using network indicators (e.g., betweenness centrality). Stirling (2007) further suggested developing a heuristics by weighing the two components; for example, by adding exponents as follows: $D = \sum_{ij(i \neq j)} (p_i \cdot p_j)^\alpha \cdot d_{ij}^\beta$.

However, one then obtains a parameter space which is infinite (Ricotta & Szeidl, 2006).[1]

In this study we limit the discussion to diversity as defined by Equation 1. The two components are the distance between the journals in the citation network ($d_{ij}$) and the variety in each journal's citation patterns. In addition to the combined formula (as provided in Equation 1), we assess potential indicators of interdisciplinarity (e.g., betweenness centrality) in the aggregated citation matrix among the journals, and the 8,207 (cited and citing) vectors using the Gini coefficient and the Shannon entropy as potential measures of interdisciplinarity.

In our opinion, it is timely to study whether this Rao-Stirling measure (Equation 1) can be elaborated into an indicator of interdisciplinarity at the journal level using the information contained in the matrix of aggregated journal-journal citation relations (cf. Rafols *et al.*, in press). Stirling (2007) proposed his approach as "a general framework for analyzing diversity in science, technology and society" because the two dimensions—(un)evenness in the distributions at the vector level and similarity among the vectors at the matrix level—are combined (Rafols & Meyer, 2010). The first part of Equation 1 (that is, $\sum_{ij(i \neq j)} p_i \cdot p_j$) is equal to the Gini-Simpson diversity measure (in biology) or the Herfindahl-Hirschman index (in economics). The Gini coefficient is akin to this measure (Atkinson, 1970; Stirling, 2007, at p. 709) and has the advantage of having been widely used in bibliometrics (e.g., Bornmann *et al*., 2008; Burrell, 1991; Cole *et al*., 1978; Danell, 2000; Frame *et al*., 1977; Persson & Melin, 1996; Rousseau, 1992, 2001; Stiftel *et al*., 2004; Zitt *et al*., 1999). We chose to compare this measure at the level of the vector of each journal's citation distributions (cited and citing) with the Shannon entropy as another indicator used in bibliometric applications (e.g., Grupp, 1990; Leydesdorff, 1991).

---

[1] Stirling (2007, at p. 712) also suggests to use only the lower triangle of the (symmetrical) matrix. Rao (1982, at pp. 6f.), however, uses the full matrix and even envisages that the main diagonal of the distance matrix may be unequal to zero depending on the research question. We used the full matrix; in our case the main diagonal values are always set to zero.

The second part of Equation 1 ( $\sum_{ij(i \neq j)} d_{ij}$ ) represents a distance matrix in a network. We will use Euclidean distances and (1 – *cosine*) as two distance measures for reasons to be explained in the methods section below. Table 2 provides an overview of the envisaged comparisons.

|  | |
|---|---|
| *Vector based* | Shannon entropy |
| | Gini index |
| *Matrix based* | Betweenness centrality |
| | Betweenness centrality (normalized) |
| *Distance measures for diversity* | Euclidean distance |
| | (1 – *cosine*) |

**Table 1**: Aggregated journal-journal citation matrix with citing and cited vectors for each journal, and six possible measures of "interdisciplinarity".

This evaluation using these six measures will be performed examining both the cited and citing dimensions of the matrix. In the (later) discussion, we focus on the cited dimension because in this dimension one also can compare with other journal indicators such as the impact factor and total cites (Bensman, 2007; Leydesdorff, 2009). Since there is as yet no established baseline for the interdisciplinarity of journals, we have to manoeuvre using correlation and factor analysis in order to understand whether two indicators measure something differently or capture to a large extent the same variation. Our initial hope was that the new Rao-Stirling diversity measure might provide us with such an obvious indicator of interdisciplinarity that one would be able to benchmark some of the other indicators against this dimension.

**Methods and materials**

*Data processing*

Data was harvested from the CD-Rom versions of the *JCRs* 2008 of the *Science Citation Index* (6,598 journals) and the *Social Sciences Citation Index* (1,980 journals). 371 of these journals are covered by both databases. Our set is therefore 6,598 + 1,980 – 371 = 8,207 journals. An asymmetrical citation matrix of these 8,207 journals "cited" versus the same journals "citing" was saved as a systems file in SPSS (v. 15). Using SPSS, one can generate a host of similarity and distance matrices. Betweenness centrality values were generated using Pajek[2] and UCINet.

Of the 8,207 journals, 8,159 are processed by the Institute of Scientific Information of Thomson Reuters (ISI) in the cited dimension and 8,064 citing. The data was collected from the citing side of the database. While single citation relations are sometimes included on the cited side, these are aggregated by the ISI under the category "All other" on the citing side. This reduction of single occurrences in the aggregated set is both computationally efficient and conceptually attractive (because incidental noise is removed).

---
[2] Pajek is freely available for academic use at http://vlado.fmf.uni-lj.si/pub/networks/pajek/.

A number of journals have thereafter no values other than on the main diagonal (that is, "journal self-citations"): 64 in the cited dimension and 20 citing. Although these journals cannot be provided with a value for the Rao-Stirling diversity indicator (since $\sum_{ij(i \neq j)} d_{ij} = 0$), we have included the diagonal values in the computation of the Gini, the probabilistic entropy, and the centrality measures.

*Gini coefficients and Shannon entropy*

Let us first turn to the vector-based measures. These are based on the frequency distributions of citations of each of the journals, either in the cited or citing directions. In the extreme case where a journal only cites or is cited by articles in the journal itself, the inequality in the citation distribution is maximal and the uncertainty minimal.[3] Maximum inequality corresponds to a Gini of unity and minimum uncertainty is equal to a Shannon entropy of zero. The journal is then extremely mono-disciplinary. In the opposite case, one would expect lower values for the Gini coefficient and higher entropy values.

---

[3] If one drew a citation, one would know for certain which journal is involved.

The Gini coefficient is a well established measure of inequality or unevenness in a distribution; it can be formulated as follows (e.g., Buchan, 2002):

$$G = \frac{\sum_{i=1}^{n}(2i-n-1)x_i}{n\sum_{i=1}^{n}x_i} \tag{3}$$

with $n$ being the number of elements in the population and $x_i$ being the number of citations of element $i$ in the ranking. The Gini ranges between zero for a completely even distribution and $(n-1)/n$ for a completely uneven distribution, approaching one for large populations. For comparisons among smaller populations of varying size, this requires a normalization that brings Gini coefficients for all populations to the same maximum unity, i.e., one. The formula for this normalised Gini coefficient is:

$$G_N = \frac{n}{(n-1)}\frac{\sum_{i=1}^{n}(2i-n-1)x_i}{n\sum_{i=1}^{n}x_i} = \frac{\sum_{i=1}^{n}(2i-n-1)x_i}{(n-1)\sum_{i=1}^{n}x_i} \tag{4}$$

The uncertainty contained in a distribution can be formalized using Shannon's (1948) formula for probabilistic entropy:

$$H = -\sum_i p_i \log(p_i) \tag{5}$$

The maximum information is the same and thus a constant for all vectors, namely $\log_2(8{,}207) = 13.00$ bits. (When the two-base of the logarithm is used, uncertainty is expressed in bits of information.) Entropy can therefore be used as an indicator without the need for further normalization. However, the entropy is aggregated for each vector from the *n* cells with only non-zero values ($n \leq N$; $N = 8{,}207$); *n* is a variable and $\log(n)$ can be considered as a local—journal-specific—maximum of the entropy. In a later section, the observed entropy values will also be assessed as a percentage of this local maximum of the entropy.

*Betweenness centrality*

Betweenness centrality is a network indicator and will be computed on both the (asymmetrical) citation matrix in both directions and the *cosine*-normalized matrices. After cosine-normalization the matrix is symmetrical, but one can normalize both over the cited and citing axis of the asymmetrical citation matrix. Thus, four evaluations are possible.

Betweenness centrality is defined as follows:

$$\sum_i \sum_j \frac{g_{ikj}}{g_{ij}}, \quad i \neq j \neq k \tag{6}$$

Or, in words: the betweenness centrality of a vertex *k* is equal to the proportion of all geodesics between pairs ($g_{ij}$) of vertices that include this vertex ($g_{ijk}$; e.g., De Nooy *et al.*, 2005, at p. 131).

In order to compute the (Freeman) betweenness centrality, the matrix is first binarized.[4] Whether a *cosine* value is larger than zero or not depends *only* on the number of co-occurrences in the numerator of the formula for the *cosine*:

$$\cos(x, y) = \frac{\sum_{i=1}^{n} x_i y_i}{\sqrt{\sum_{i=1}^{n} x_i^2} \sqrt{\sum_{i=1}^{n} y_i^2}} = \frac{\sum_{i=1}^{n} x_i y_i}{\sqrt{(\sum_{i=1}^{n} x_i^2) * (\sum_{i=1}^{n} y_i^2)}} \tag{7}$$

In other words, betweenness centrality in the *cosine*-normalized vector space is after binarization for analytical reasons equal to betweenness centrality in the co-occurrence matrices (cited or citing).[5] This inference is only valid when using no threshold values for the *cosine*. In this study, however, we did not set a threshold because a threshold introduces another parameter into the model and would thus complicate the comparison.

The advantage of using co-occurrence—in our case, co-citation—matrices instead of *cosine*-normalized matrices is computational given the large size of the matrix: the (1-

---

[4] In the meantime, algorithms for weighting the line values are available (Brandes, 2001, 2008), but the implementation in Visone (at http://www.visone.info) cannot (yet) handle data matrices of this size (Brandes, *personal communication*, 25 February 2010).
[5] Using SPSS (v. 15) for the construction of *cosine* matrices and Ucinet (v. 6.267) for the affiliations matrices (both cited and citing), we found correlations between the betweenness values of $r = 0.987$ (cited) and $r = 0.990$ (citing). These values should be precisely 1.0. We cannot explain these differences other than as an effect of potentially different roundings of decimals.

mode) co-citation matrix can be obtained from the asymmetrical (2-mode) citation matrix by multiplication with the transposed of the latter matrix. This can be done both in the cited ($AA^T$) and the citing dimension ($A^TA$) of the asymmetrical citation matrix. We shall thus compare betweenness centrality as an indicator of interdisciplinarity by using both the asymmetrical citation matrix (in both the cited and citing directions) and the two symmetrical co-citation matrices (that is, using the numerators of Equation 7 for distinguishing between zeros and ones).

*Distance matrices (Rao-Stirling diversity)*

Euclidean distances seem a most natural candidate for the distance matrix used for measuring Rao-Stirling diversity (Equation 1). First, Euclidean distances involve the least restrictive assumptions; second, Euclidean distances can be transformed through simple scaling of dimensions to represent a wide range of possible geometries (Kruskal, 1964); and third, Euclidian distances are more familiar, parsimonious, and intuitively accessible than most other distance measures. However, in our case, two journals with precisely the same distributions but with different sizes would be counted as distant from each other. For this reason, one first needs to normalize the data for size.

In order to prevent this size effect, one could consider using *z*-scores (Bornmann & Daniel, 2009). However, Ahlgren *et al*. (2003) have shown that the Pearson correlation can be heavily affected by the large number of zeros in citation matrices; they therefore proposed the *cosine* as a non-parametric similarity criterion. Analogously to the Pearson correlation, *z*-scores are based on averages and equally affected by this effect of the zeros on the mean. The *cosine*, however, is a similarity measure and not a distance. Yet, (1 – *cosine*) represents a dissimilarity, and thus can be considered as a relevant measure of distance.

Alternatively, using Euclidean distances one can prevent the noted effect of differences in size in otherwise similar vectors (rows or columns) by using the proportions of citations, that is, the relative instead of absolute frequency distributions. (Conveniently, the relative frequency distribution is by definition equal to the probability distribution that we need for computing the other part of the Rao-Stirling diversity.) The noted effect of size is neutralized by this normalization. In sum, we will perform the computation of Rao-Stirling diversity using the four possibilities of these normalized Euclidean distances and (1 – *cosine*) versus cited and citing.

**Results**

Throughout the paper, we use Spearman's rank-order correlations because our primary objective is an indication of interdisciplinarity as a variable attribute among journals. Some of the comparisons among specific indicators (e.g., before and after normalization) allow for finer-grained comparisons using the Pearson correlation, but in order to prevent confusion we use the Spearman correlation consistently.

*a. Gini coefficients and the Shannon entropy*

Let us first consider Gini coefficients and Shannon entropy as two vector-based journal indicators, both in the cited and the citing directions of the citation matrix. While the Gini coefficient indicates unevenness, Shannon entropy provides an indicator of evenness. In other words, the Gini coefficient can be considered as an indicator of specificity and therefore disciplinarity, whereas the entropy ($H$) increases both when more cells of the vector are affected and with greater spread among the different categories. Thus, this indicator can be expected to capture both the aspects of a wider range and a more equally distributed set of citing or cited journals, respectively. Therefore, the entropy measure can be expected to be sensitive to size, whereas the Gini coefficient is normalized in the denominator of Equation 4.

|   |   | Gini cited | entropy cited | Gini citing | entropy citing |
|---|---|---|---|---|---|
| Gini cited | Correlation Coefficient | 1.000 | -.803(**) | .423(**) | -.449(**) |
|  | Sig. (2-tailed) | . | .000 | .000 | .000 |
|  | N | 8,207 | 8,207 | 8,207 | 8,207 |
| entropy cited | Correlation Coefficient | **-.803**(**) | 1.000 | -.351(**) | .631(**) |
|  | Sig. (2-tailed) | .000 | . | .000 | .000 |
|  | N | 8,207 | 8,207 | 8,207 | 8,207 |
| Gini citing | Correlation Coefficient | .423(**) | -.351(**) | 1.000 | -.658(**) |
|  | Sig. (2-tailed) | .000 | .000 | . | .000 |
|  | N | 8,207 | 8,207 | 8,207 | 8,207 |
| entropy citing | Correlation Coefficient | -.449(**) | .631(**) | **-.658**(**) | 1.000 |
|  | Sig. (2-tailed) | .000 | .000 | .000 | . |
|  | N | 8,207 | 8,207 | 8,207 | 8,207 |

** Correlation is significant at the 0.01 level (2-tailed).

**Table 2**: Spearman's $\rho$ among Gini coefficients and Shannon entropies; both cited and citing ($N = 8,207$ journals).

The negative signs of the rank-order correlations between the two indicators (Table 2) show the opposite directionality. Not surprisingly, there is no strong correlation between rankings in the cited and citing dimensions: journals that build on diverse knowledge bases (citing patterns) do not necessarily have diverse audiences (cited patterns). Table 2 shows that correlations between the two indicators in the cited dimension ($\rho = -0.803$) are higher than in the citing dimension ($\rho = -0.658$).[6] This is understandable, since the citing side represents the research front and therefore introduces variability, while the archive of science is cited and thus can be expected to be more stable (Leydesdorff, 1993). One can expect these indicators to show such differences in functionality between cited and citing.

Table 3 shows the top 20 journals in terms of Gini coefficients (after correction for the ascending and descending orders) and entropy, both cited and citing. As expected, the entropy measure is affected by size. The listed journals are recognizable as intuitively the most interdisciplinary among the journals (in both the cited and citing dimensions; columns b and d, respectively). The perception of the importance of journals is also heavily influenced by their size (Bensman, 2007), so this intuitive recognition is not incidental (cf. Bollen *et al.*, 2009a, at p. 6).

---

[6] All correlations are highly significant because of the large number of cases ($N = 8,207$).

| *Gini; cited* (a) | *entropy; cited* (b) | *Gini; citing* (c) | *entropy; citing* (d) |
|---|---|---|---|
| Brit Med Bull | Science | Braz J Med Biol Res | Ann NY Acad Sci |
| Sci Am | Lancet | Yonsei Med J | Med Hypotheses |
| New Engl J Med | Nature | J Zhejiang Univ-Sc A | Curr Pharm Design |
| Lab Invest | Ann NY Acad Sci | J Zhejiang Univ-Sc B | Afr J Biotechnol |
| Cell Mol Biol | New Engl J Med | Afr J Biotechnol | Curr Med Chem |
| Annu Rev Med | P Natl Acad Sci USA | Drugs R&D | Cochrane Db Syst Rev |
| J Lab Clin Med | Jama-J Am Med Assoc | Toxicol Mech Method | Saudi Med J |
| Am J Med | Brit Med J | Int J Exp Pathol | Med Sci Monitor |
| Psychol Bull | Biochem Bioph Res Co | Biomed Environ Sci | Curr Sci India |
| Med Clin N Am | Am J Med | Ital J Zool | Expert Opin Pharmaco |
| Ann Med | Ann Intern Med | Adv Ther | Int J Mol Sci |
| Clin Sci | Psychol Bull | Eur J Med Res | Chinese Med J-Peking |
| Annu Rev Psychol | Faseb J | J Korean Med Sci | Front Biosci |
| Jama-J Am Med Assoc | J Clin Invest | Acta Med Okayama | Mini-Rev Med Chem |
| Eur J Clin Invest | Nat Med | J Int Med Res | Biol Pharm Bull |
| Ann NY Acad Sci | Am J Pathol | Arab J Sci Eng | Chinese Sci Bull |
| Nat Biotechnol | Arch Intern Med | J Formos Med Assoc | J Int Med Res |
| J Intern Med | Anal Biochem | Arch Psychiat Nurs | Exp Biol Med |
| Qjm-Int J Med | Febs Lett | Orphanet J Rare Dis | Yonsei Med J |
| Postgrad Med J | Biochem J | Fund Clin Pharmacol | Braz J Med Biol Res |

**Table 3**: The top 20 journals in terms of Gini coefficients and Shannon entropy; both cited and citing ($N = 8{,}207$).[7]

The Gini coefficient corrects for this size effect because of a normalization in the denominator. In the citing dimension (column c), this measure provides us with a number of peripheral journals exhibiting the highest interdisciplinarity in aggregated referencing behavior. In other words, knowledge from very different domains is cited within these journals. In a study of aggregated citations among Chinese journals, Leydesdorff & Jin (2005) have noted that university journals in China often have the specific function of combining different knowledge bases from a sectorial or institutional perspective. On the cited side, *Scientific American* (0.0102) emerges as the second most highly ranked journal, surpassed only by the *British Medical Bulletin* (0.0097).

---

[7] Because of its different sign, the Gini is sorted ascendingly. This brings 48 journals (cited) and 144 journals (citing) to the top of the rank-ordering with a Gini of zero, but these journals were not included in these lists.

*b. Betweenness centrality*

As noted, Leydesdorff (2007) recommended using betweenness centrality in the *cosine*-normalized vector space as another indicator of interdisciplinarity. *Cosine*-normalization was proposed because before normalization of betweenness centrality and degree centrality can be correlated,[8] and thus defined, interdisciplinarity would partly be a reflection of size. We shall now compare the four options for computing betweenness centrality in the asymmetrical citation matrix and the cosine-normalized matrices in both directions (cited and citing). No threshold will be used because citation densities can be expected to vary among fields of science (Moed, 2010).[9]

|  |  | Citations cited | Cosine cited | Citations citing | Cosine citing |
|---|---|---|---|---|---|
| Citations (cited) | Correlation Coefficient | 1.000 | .915(**) | .602(**) | .488(**) |
|  | Sig. (2-tailed) | . | .000 | .000 | .000 |
|  | N | 8,207 | 8,207 | 8,207 | 8,207 |
| Cosine (cited) | Correlation Coefficient | **.915(**)** | 1.000 | .627(**) | .597(**) |
|  | Sig. (2-tailed) | .000 | . | .000 | .000 |
|  | N | 8,207 | 8,207 | 8,207 | 8,207 |
| Citations (citing) | Correlation Coefficient | .602(**) | .627(**) | 1.000 | .741(**) |
|  | Sig. (2-tailed) | .000 | .000 | . | .000 |
|  | N | 8,207 | 8,207 | 8,207 | 8,207 |
| Cosine (citing) | Correlation Coefficient | .488(**) | .597(**) | **.741(**)** | 1.000 |
|  | Sig. (2-tailed) | .000 | .000 | .000 | . |
|  | N | 8,207 | 8,207 | 8,207 | 8,207 |

\*\* Correlation is significant at the 0.01 level (2-tailed).

**Table 4**: Spearman's $\rho$ among betweenness centralities in the asymmetrical citation matrix versus cosine-normalized matrices; both cited and citing ($N$ = 8,207 journals).

---

[8] The Pearson correlations between these two centrality measures in this journal set ($N$ = 8,207) are $r$ = 0.508 (cited) and 0.566 (citing); Spearman's $\rho$ = 0.877 and 0.777, respectively.
[9] Leydesdorff (2007) used *cosine* $\geq$ 0.2 to study relatively limited and therefore more homogenous fields.

Again, we find higher correlations on the cited side than citing (Table 4). By listing the top 20 journals in each ranking (Table 5), we are able to discern more details. The first two columns (a and b) of Table 5 show the predicted effect of size. Without normalization *Science*, *Nature*, and *PNAS* are indicated as the major interdisciplinary journals.[10] On the citing side (third column), however, these three journals follow at the 10$^{th}$, 16$^{th}$, and 1$^{st}$ positions, respectively. Normalization (using the *cosine*) changes this to the extent that none of these three journals is in the top-20 list citing (column d). In the cited dimension (column b), other mainly social science journals are listed as most interdisciplinary.

| *Citations; cited* (a) | *Cosine-normalized; cited* (b) | *Citations; citing* (c) | *Cosine-normalized; citing* (d) |
|---|---|---|---|
| Science | Econometrica | P Natl Acad Sci USA | Ecol Econ |
| Nature | Psychol Bull | Ecol Econ | Energ Policy |
| P Natl Acad Sci USA | Manage Sci | Ann NY Acad Sci | Am J Public Health |
| Psychol Bull | Psychol Rev | Am J Public Health | J Am Soc Inf Sci Tec |
| Econometrica | Am J Public Health | J Am Soc Inf Sci Tec | Risk Anal |
| Am J Public Health | Science | Phys Rev E | Global Environ Chang |
| Psychol Rev | J Econometrics | Energ Policy | Manage Sci |
| Manage Sci | Ecol Econ | Risk Anal | Soc Stud Sci |
| Lancet | Risk Anal | J Geophys Res | Technol Cult |
| Am J Psychiat | Energ Policy | Science | Annu Rev Inform Sci |
| New Engl J Med | Behav Brain Sci | J Agr Food Chem | Psychol Bull |
| Ann NY Acad Sci | Nature | J Clin Nurs | Scientometrics |
| Jama-J Am Med Assoc | P Natl Acad Sci USA | Environ Sci Technol | J Epidemiol Commun H |
| J Biol Chem | Annu Rev Psychol | Ieee Eng Med Biol | Omega-Int J Manage S |
| Phys Rev Lett | Psychometrika | Sensors-Basel | Annu Rev Env Resour |
| Brit Med J | Am J Psychiat | Behav Brain Sci | Behav Brain Sci |
| Annu Rev Psychol | J R Stat Soc A Stat | Nature | Math Soc Sci |
| Arch Gen Psychiat | Struct Equ Modeling | Philos T R Soc B | Comput Educ |
| J Am Stat Assoc | Jama-J Am Med Assoc | Front Biosci | Technovation |
| Nucleic Acids Res | J Am Stat Assoc | Appl Math Comput | J Archaeol Sci |

**Table 5**: Betweenness centrality for the top-20 journals in the four categories based on raw versus cosine-normalized citation scores; cited versus citing ($N$ = 8,207).

---

[10] This result accords with Bollen *et al.*'s (2009b, at p. 6) findings.

On the citing side (column d), the *Journal of the American Society of Information Science & Technology,* the *Annual Review of Information Science & Technology*, and *Scientometrics* are among the lead journals together with other journals in science & technology studies. Perhaps this indicates that authors in these journals are not only making references for their discursive arguments, but also studying other sciences, from which they incidentally cite. Note that any value above zero counts fully in the binarized matrix which is used for the computation of betweenness centrality. Similarly, it is not amazing that some of the statistics journals are cited in a large range of disciplines. As noted above (in footnote 4), algorithms for weighted betweenness centrality are available in the literature (Brandes, 2001), but have not yet been implemented for matrices of this size (Brandes, *personal communication*, 25 February 2010). We return to this possible elaboration in our conclusions.

In summary, the indicator of betweenness centrality provides us with understandable results after normalization. A full listing of the 8,207 journals with the 12 indicators discussed in this study is available online at http://www.leydesdorff.net/jcr08/interdisciplinarity/rankings.htm.[11] Furthermore, one can download at http://www.leydesdorff.net/jcr08/interdisciplinarity/indicators.xls the Excel sheet and sort according to one's own preferences.

---

[11] This file is 22.7 Mbyte.

*c. Rao-Stirling diversity*

Table 6 shows the Spearman rank-correlations among the four options of computing Rao-Stirling diversity using (1 – *cosine*) or Euclidean distances versus the cited and citing dimensions of the aggregated citation matrix among the 8,207 journals.

|  |  | (1 – c*osine*); cited | Euclidean distances; cited | (1 – c*osine*); citing | Euclidean distances; citing |
|---|---|---|---|---|---|
| (1 – c*osine*); cited | Correlation Coefficient | 1.000 | -.012 | .303(**) | .254(**) |
|  | Sig. (2-tailed) | . | .290 | .000 | .000 |
|  | N | 8,207 | 8,207 | 8,207 | 8,207 |
| Eucl. distances; cited | Correlation Coefficient | **-.012** | 1.000 | .140(**) | .028(*) |
|  | Sig. (2-tailed) | .290 | . | .000 | .012 |
|  | N | 8,207 | 8,207 | 8,207 | 8,207 |
| (1 – c*osine*); citing | Correlation Coefficient | .303(**) | .140(**) | 1.000 | -.015 |
|  | Sig. (2-tailed) | .000 | .000 | . | .164 |
|  | N | 8,207 | 8,207 | 8,207 | 8,207 |
| Eucl. distances; citing | Correlation Coefficient | .254(**) | .028(*) | **-.015** | 1.000 |
|  | Sig. (2-tailed) | .000 | .012 | .164 | . |
|  | N | 8,207 | 8,207 | 8,207 | 8,207 |

\*\* Correlation is significant at the 0.01 level (2-tailed).
\* Correlation is significant at the 0.05 level (2-tailed).

**Table 6**: Rank-order correlations (Spearman's $\rho$) among diversity measures based on (1 – *cosine*) and Euclidean distances; *cited* and *citing* ($N$ = 8,207).

Not surprisingly, there is no strong correlation between rankings in the cited and citing dimensions. However, the negative (albeit not significant) correlation between the (1 – c*osine*)-based and distance-based diversities is unexpected. The resulting diversity is thus heavily dependent on the choice of the distance measure.

Let us try to assess qualitatively which of the two measures would make more sense by listing the top 20 journals in these four cases (Table 7).

| (1 – *cosine*); cited (a) | *Euclidean distances*; cited (b) | (1 – *cosine*); citing (c) | *Euclidean distances*; citing (d) |
|---|---|---|---|
| Sci Am | Zootaxa | J Zhejiang Univ-Sc A | Phys Today |
| Comput J | Rev Mex Biodivers | Simul Model Pract Th | Astron Geophys |
| Am Behav Sci | Zoosystema | Arab J Sci Eng | J Astrophys Astron |
| P Ieee | T Am Entomol Soc | J Chin Inst Eng | Nat Phys |
| Mt Sinai J Med | J Agr U Puerto Rico | Lat Am Appl Res | Space Sci Rev |
| Daedalus-Us | Deut Entomol Z | Iran J Sci Technol A | Contemp Phys |
| Curr Sci India | P Entomol Soc Wash | Iran J Sci Technol B | Rep Prog Phys |
| Am Fam Physician | Coleopts Bull | Appl Math Model | Phys Rev Lett |
| Postgrad Med | Trop Zool | Math Comput Simulat | Ann Phys-Paris |
| South Med J | Ann Soc Entomol Fr | J Mech Sci Technol | Cr Phys |
| J R Soc Med | Raffles B Zool | Expert Rev Med Devic | New J Phys |
| Yonsei Med J | Rec Aust Mus | Electr Eng Jpn | Chem Eng News |
| Postgrad Med J | Orient Insects | P I Mech Eng C-J Mec | Riv Nuovo Cimento |
| Med Clin N Am | Prof Eng | Paedagog Hist | Adv Space Res |
| Can Med Assoc J | Int J Acarol | J Cent South Univ T | Epl-Europhys Lett |
| Psychol Bull | J Hymenopt Res | Inverse Probl Sci En | Am Sci |
| Am J Med Sci | Aquat Insect | Int J Med Robot Comp | Nuovo Cimento B |
| J Postgrad Med | Space Sci Rev | Comput Meth Prog Bio | Nat Photonics |
| Brit Med J | Rev Suisse Zool | Sadhana-Acad P Eng S | Top Appl Phys |
| Am J Med | Pan-Pac Entomol | Eng Appl Artif Intel | Nat Mater |

**Table 7**: The top 20 journals in the four categories of Rao-Stirling diversity measures (based on (1 – *cosine*) versus Euclidean distances; cited versus citing).

The results based on using (1 – *cosine*) as a distance measure can be provided with an interpretation, but an interpretation is more difficult to provide for results based on Euclidean distances. That the *Scientific American* (impact factor = 2.316) is a top journal in terms of being cited interdisciplinarily is no surprise. The diversity in the citation patterns of the journals in this first column (a) of Table 2 ranges from $D = 0.465$ for the *American Journal of Medicine* to $D = 0.482$ for the *Scientific American*.

In the citing dimension (column c), the same measure (1 – *cosine*) provides us with peripheral journals as above in the case of using the Gini coefficient. For example, authors in the *Journal of the Zhejiang University–Science A* cite interdisciplinarily to the extent that $D = 0.475$. (This journal had an impact factor of 0.554 in 2008.) The two other columns (b and d) are based on Euclidean distances; cited and citing, respectively. In the cited dimension, zoology journals seem to prevail, and physics journals in the citing dimension. These results are counter-intuitive.

Garfield (1972, at p. 478) mentioned ecology as a field in which one cited from very different knowledge bases, but was at that time not cited from outside this domain. Thus, this field operated as a sink of citations. Garfield's exercise was based on data for the final quarter of 1969. In 2008, the journal *Ecology* contained 40,749 citations with an indegree of 948 (234[th] position) as against 18,752 references with an outdegree of 482 (648[th] position).

The journal *Ecology* occupies position 2,694 in the ranking of citing interdisciplinarily based on Euclidean distances ($D = 0.320$). The same journal ranks even lower (in the 6,149[th] place) using the (1 – *cosine*)-based distance matrix ($D = 0.296$). In the cited dimension, this journal is ranked at the 1,830[th] place ($D = 0.407$), and at the 4,468[th] place using Euclidean distances ($D = 0.195$). Obviously, most citations and references of this journal are from nearby vertices in the network, so that the distances are not large. The journal therefore cannot be considered as high in interdisciplinarity, neither in its cited or citing behavior, according to these indicators.

In summary, these results first suggest that the (1 – *cosine*)-based measure operates on average better as an indicator of interdisciplinarity than the one based on Euclidean distances. However, not all the results are convincing. For example, the relatively low positions of *Nature* (191[st] with $D$ = 0.447) and *Science* (169[th] with $D$ = 0.448) in the ranking can perhaps be explained because these journals and *PNAS* (635[th] position with $D$ = 0.430) are cited in common patterns. Rafols & Meyer (2010, at pp. 275f.) found a high percentage (> 33%) of references to these three journals in bio-nano publications. Nevertheless, these scores remain unconvincing from a bibliometric perspective.

|  | (1 – *cosine*); cited | Euclidean distances; cited | (1 – *cosine*); citing | Euclidean distances; citing |
|---|---|---|---|---|
| Mean | .354939094 | .207695013 | .323983699 | .292496223 |
| Std. Deviation | .0776163635 | .0589444533 | .0737042333 | .0804055900 |
| Variance | .006 | .003 | .005 | .006 |
| Range | .4822650 | .5288082 | .4748092 | .6563784 |

**Table 8**: Statistics of the means, standard deviations, variances, and ranges in the case of diversity measures based on *cosine* values or Euclidean distances and cited versus citing ($N$ = 8,207).

Another problem is the discriminating power of the indicator, with differences sometimes only in the third decimal. Table 8 provides some statistics about the mean standard deviation, the variance, and the range (from zero). The multiplication of two indicators both ranging between zero and one, of course, depresses the outcome. (Perhaps one should consider vector-based measures quantitatively with the square root of Rao-Stirling diversity.)

In summary, this journey through three possible types of indicators of interdisciplinarity—vector-based, matrix-based, or combined—leads us to conclude that the different indicators provide insights about different aspects of interdisciplinarity. Shannon entropy measures variety at the vector level and can be thus used as an indicator of interdisciplinarity if one is not primarily interested in a correction for size-effects. Betweenness centrality in the *cosine*-normalized matrix provides a measure for interdisciplinarity. Using cosine *values* as weights for the edges can be expected to improve this measure further.[12] Rao-Stirling diversity measures are sensitive to the distance measure being used. The possibility to weigh the two parts of the Equation 1 with different exponents would complicate the identification of an indicator for interdisciplinarity: in an infinite parameter space, one can always find a best fit. This would further complicate the identification of a simple and robust indicator of interdisciplinarity.

**Relations among the various indicators**

Factor analysis enables us to study whether the various indicators cover the same ground or should be considered as different. Leydesdorff (2009) found two main dimensions—namely, size and impact—in the cited direction when using the ISI set of journals and including network indicators. On the one hand, the impact factor and the immediacy index are highly correlated (Yue *et al*., 2004); on the other, total cites and indegree can be considered as indicators of size (Bensman, 2007; Bollen *et al*., 2009a and b).

---

[12] Bollen *et al*. (2009) found hardly any difference between weighted and unweighted betweenness centrality using non-normalized data.

|  | Component | | |
|---|---|---|---|
|  | 1 | 2 | 3 |
| Betweenness (citations) | .881 | | |
| Total cites | .854 | | .156 |
| Indegree | .721 | .387 | .379 |
| Betweenness (*cosine*-normalized) | .719 | .291 | |
| Shannon entropy | .257 | .896 | .208 |
| Gini coefficient | | -.892 | |
| Rao-Stirling (1 – *cosine*) | | .890 | |
| Immediacy | .215 | .125 | .780 |
| Impact factor | .351 | .265 | .752 |
| Rao-Stirling (Euclidean) | .111 | .197 | *-.531* |

Extraction Method: Principal Component Analysis.
 Rotation Method: Varimax with Kaiser Normalization.
a  Rotation converged in 4 iterations.

**Table 9**: Rotated Component Matrix of indicators in the cited dimension

Using these four indicators to anchor the two main dimensions in the *cited* dimension and the six indicators discussed above, Table 9 shows that in a three-factor model—three factors explain 72.4% of the variance in this case—the first factor can indeed be associated with "size" and the third with "impact." Entropy, the Gini coefficient, and Rao-Stirling diversity based on (1 – *cosine*) as a distance measure constitute another (second) dimension which one could designate as "interdisciplinarity." Betweenness centrality, however, loads highest on the size factor even after normalization for size. Rao-Stirling diversity based on relative Euclidean distances loads negatively on the third factor ("impact"), and is in this respect different from all the other indicators under study. As noted, we could not provide this variant of the indicator with an interpretation.

If we focus—in a second model—on the six variables that were studied above (diversity based on two different distance matrices, between centrality before and after normalization, the Gini coefficient, and probabilistic entropy) we can explore their internal structure using factor analysis in both the cited and citing directions.

| *Cited* | Component | | |
|---|---|---|---|
| | 1 | 2 | 3 |
| Shannon entropy | .921 | .231 | |
| Gini coefficient | -.899 | | |
| Rao-Stirling (1 – *cosine*) | .872 | | .210 |
| Betweenness (citations) | | .903 | |
| Betweenness (*cosine*-normalized) | .283 | .838 | |
| Rao-Stirling (Euclidean distances) | | | .991 |

Extraction Method: Principal Component Analysis.
 Rotation Method: Varimax with Kaiser Normalization.
a  Rotation converged in 4 iterations.

| *Citing* | Component | | |
|---|---|---|---|
| | 1 | 2 | 3 |
| Rao-Stirling (Euclidean distances) | .860 | | -.111 |
| Rao-Stirling (1 – *cosine*) | .673 | | .331 |
| Shannon entropy | .663 | .208 | .619 |
| Betweenness (citations) | | .897 | |
| Betweenness (*cosine*-normalized) | .116 | .880 | |
| Gini coefficient | | | -.938 |

Extraction Method: Principal Component Analysis.
 Rotation Method: Varimax with Kaiser Normalization.
a  Rotation converged in 4 iterations.

**Table 10a and b**: Three-factor solution for the six indicators in the cited and citing direction, respectively. Three factors explain 85.2 and 78.1% of the variance, respectively; $N = 8,207$.

The factor structures in Tables 10 (a and b)—cited and citing, respectively—are considerably different. These results suggest that the underlying structure is more determined by the functionality in the data matrix (cited or citing) than by correlations among the indicators. In both solutions, however, betweenness before and after normalization load together on a second factor. This is not surprising since the two measures are related (Bollen *et al.*, 2009b). In both solutions, we also find Rao-Stirling diversity measured on the basis of (1 – *cosine*) as the distance measure and Shannon entropy loading on the same factor. The Gini coefficient and the Rao-Stirling diversity based on Euclidean distances have a different (i.e., not consistent) position in the cited or the citing directions.

In summary, Shannon entropy qualifies as a vector-based measure of interdisciplinarity. Our assumption that the Gini coefficient would qualify as an indicator of inequality and therefore (disciplinary) specificity was erroneous: interdisciplinarity is not just the opposite of disciplinarity. Betweenness centrality and Rao-Stirling diversity (after cosine-normalizations) indicate different aspects of interdisciplinarity. Betweenness centrality, however, remains associated with size more than Rao-Stirling diversity or entropy despite the normalization. Perhaps setting a threshold would change this dependency on size because larger journals can be expected to be cited in or citing from a larger set.

**Library and information science**

Because of the explorative nature of this study, we thought it appropriate to zoom in on one of the subject categories of the ISI despite their well-known shortcomings (Boyack *et al.*, 2005; Rafols & Leydesdorff, 2009). In order to enhance the interpretation by the readership of this journal, we chose the category of library and information science, which contained 61 journals in 2008. In other words, we compare these 61 journals in terms of how they are cited by the 8,207 journals in the database. (Note that one can also compute local values for betweenness centrality, etc., using the 61 x 61 citation matrix among these journals.)

As expected, the correlations are somewhat different for this subset of 61 journals from the overall set of 8,207 journals. Notably, the Rao-Stirling diversity indices measured on the basis of (1 – *cosine*) and relative Euclidean distances are no longer correlated negatively, but $\rho = 0.230$. More interestingly, the factor structures, again using a three-factor solution and Varimax rotation, are now comparable between the cited and citing dimensions, and both are similar to the structure in the citing case in the previous section. By focusing on these 61 journals, we have removed outliers (such as *Nature* and *Science*) on the one hand and peripheral journals (for example, journals with no citations other than self-citations) on the other. Therefore, we consider these structures to be reliable.

| *Cited* | Component | | |
|---|---|---|---|
| | 1 | 2 | 3 |
| Rao-Stirling (Euclidean distances) | .880 | .148 | .134 |
| Rao-Stirling (1 – *cosine*) | .876 | | .316 |
| Shannon entropy | .790 | .267 | .469 |
| Betweenness (citations) | .118 | .977 | |
| Betweenness (*cosine*-normalized) | .147 | .967 | .120 |
| Gini coefficient | -.400 | -.104 | -.895 |

Extraction Method: Principal Component Analysis.
 Rotation Method: Varimax with Kaiser Normalization.
a  Rotation converged in 4 iterations.

| *Citing* | Component | | |
|---|---|---|---|
| | 1 | 2 | 3 |
| Rao-Stirling (1 – *cosine*) | .920 | | -.182 |
| Rao-Stirling (Euclidean distances) | .856 | .185 | -.176 |
| Shannon entropy | .825 | .284 | -.387 |
| Betweenness (citations) | .102 | .952 | |
| Betweenness (*cosine*-normalized) | .255 | .909 | -.126 |
| Gini coefficient | -.336 | | .935 |

Extraction Method: Principal Component Analysis.
 Rotation Method: Varimax with Kaiser Normalization.
a  Rotation converged in 4 iterations.

**Table 11a and b**: Three-factor solution for the six indicators in the cited and citing direction, respectively, for 61 journals subsumed under the ISI subject category "library and information science." Three factors explain 91.9% and 90.3% of the variance, respectively.

The two factor structures provided in Table 11 (a and b) show first that the Gini coefficient measures something specific. With hindsight, we should perhaps not have considered this measure as a serious measure of interdisciplinarity. The two ways to measure betweenness centrality and Rao-Stirling diversity, respectively, provide the first two factors. Entropy loads primarily on Factor One with Rao-Stirling diversity, and to a lower extent on Factor Two with betweenness centrality. Perhaps one could envisage another indicator in which one would feed the contributions to the Shannon entropy (that is, $-p_i \log p_i$) instead of $p_i$ into Equation 1; however, this would lead us beyond the scope of the present study (Hill, 1973; Ricotta & Szeidl, 2006; Stirling, 1998, at p. 49f.).

|  | Rao-Stirling (1 – *cosine*) | Rao-Stirling (Euclidean) | Betweenness (citations) | Betweenness (*cosine*-normalized) | Gini coefficient | Shannon entropy |
|---|---|---|---|---|---|---|
| J Am Soc Inf Sci Tec | 33 | 1 | 1 | **1** | 18 | 8 |
| Scientometrics | 48 | 9 | 2 | **2** | 33 | 24 |
| Int J Geogr Inf Sci | 2 | 32 | 4 | **3** | 17 | 5 |
| Mis Quart | 10 | 38 | 3 | **4** | 3 | 1 |
| Inform Manage-Amster | 5 | 28 | 6 | **5** | 8 | 3 |
| J Am Med Inform Assn | 4 | 51 | 7 | **6** | 22 | 4 |
| J Manage Inform Syst | 13 | 35 | 8 | **7** | 5 | 2 |
| Inform Process Manag | 31 | 4 | 5 | **8** | 19 | 9 |
| J Inf Sci | 41 | 6 | 11 | **9** | 28 | 16 |
| Soc Sci Comput Rev | 7 | 24 | 12 | **10** | 26 | 13 |
| Scientist | 20 | 15 | 21 | **11** | 32 | 29 |
| J Inf Technol | 9 | 29 | 14 | **12** | 4 | 6 |
| Telecommun Policy | 37 | 47 | 15 | **13** | 43 | 38 |
| Inform Syst Res | 12 | 36 | 18 | **14** | 7 | 7 |
| J Health Commun | 11 | 49 | 29 | **15** | 16 | 10 |
| J Comput-Mediat Comm | 8 | 30 | 16 | **16** | 39 | 12 |
| Annu Rev Inform Sci | 52 | 3 | 20 | **17** | 12 | 19 |
| Online Inform Rev | 29 | 18 | 13 | **18** | 35 | 30 |
| Inform Soc | 3 | 21 | 30 | **19** | 11 | 11 |
| Soc Sci Inform | 1 | 41 | 34 | **20** | 24 | 14 |
| (…) | … | … | … | … | … | … |
| J Informetr | **51** | **40** | **48** | **51** | **49** | **51** |

**Table 12**: Top 20 journals in the ISI-category Library and Information Science sorted on betweenness centrality in the being-cited patterns after normalization. The numbers express the rank. The *Journal of Informetrics* was added at the bottom of the list.

Table 12, finally, shows the top 20 journals ranked on their betweenness centrality after normalization as one of the possible indicators for interdisciplinarity. Entropy correlates at the level of $\rho = 0.830$ with this indicator, and $\rho = 0.732$ with Rao-Stirling diversity based on $(1 - cosine)$ as the distance measure. The latter measure has the advantage of correlating less with size (for example, total cites) than the other two: the $\rho$ with total cites (in 2008) was 0.549 for Rao-Stirling diversity, 0.880 for betweenness centrality, and 0.793 for Shannon entropy.

Although the rankings differ considerably at the top, we added the *Journal of Informetrics* as a lower-ranked journal. With the exception of Rao-Stirling diversity measured on the basis of Euclidean distances—which we found not to be a good indicator, as shown above—all other indicators rank this journal between 48 and 51 on a list of 61. The journal therefore is ranked as disciplinary, and is in this respect very different from the *JASIST* or *Scientometrics*.

**Conclusions and discussion**

We did not find a robust and unambiguous indicator of "interdisciplinarity" that would be welcome from a policy perspective. However, let us keep in mind that the ISI-impact factor is also not by definition the best and only indicator of impact, but only by convention and for reasons of convenience (Leydesdorff & Bornmann, *in preparation*; Moed, 2010; Zitt, 2010). The three classes of indicators investigated here can all be considered possible candidates for an indicator of interdisciplinarity.

Among the vector-based indicators, the Shannon entropy takes into account both the reach of a journal in terms of its degree—because this number ($n \leq N$; $N = 8,207$) limits the maximal uncertainty within the subset—and the spread in the pattern of citations among these *n* journals. By normalizing the entropy as a percentage of this local maximum ($\log(n)$), one can correct for the size effect. But this brings to the top of the ranking specialist journals that are cited equally across a relatively small set. For example, there are 16 journals in the set with an entropy that is 100% of this maximum entropy; among these are the *Journal of Nanomaterials* cited by articles in six journals and *Brain Cell Biology* cited only by two journals.[13] Thus, there is no easy way to correct for the size effect by using entropy. Nevertheless, the results are intuitively easy to understand and the entropy is straightforward in the computation.

Betweenness centrality can be computed by using one of the programs for social network analysis. We derived that the betweenness centrality in the co-citation matrix is equal to that in the *cosine*-normalized matrix because of the binarization involved. This facilitates the computation (using Pajek, UCINet, or any other such program). However, the effects of the normalization were smaller than we expected in quantitative terms. Small effects may be meaningful in the ranking, as could be seen by comparing columns a and b of Table 5: whereas in column (a) *Science* and *Nature* are at the top of the list, they rank only at the 6th and 12th positions using the cosine-normalized matrix. Betweenness centrality based on cosine-normalized matrices qualifies as an indicator of interdisciplinarity. Note that betweenness in the vector space (normalized by the cosine) is positional instead of relational (Burt, 1982; Leydesdorff, 2007).

---

[13] As noted in the methods section, single citation relations were excluded from the analysis because the data was processed in terms of the citing dimension.

This study was triggered by the idea that Stirling's (2007) heuristics of diversity would provide us with sophisticated indicators of the interdisciplinarity of journals (Rafols & Meyer, 2010). The indicator is sophisticated in that it combines the distance in a network with the distribution in each vector. By setting the exponents for these two factors in the equation to zero, respectively, one can understand the relative contributions of the unevenness in the vectors and the similarities in the matrix. Higher exponents are also possible, as Stirling (2007) argues, but this would lead us beyond the scope of this study.

The results of using the Rao-Stirling diversity measure were disappointing. The measure is very sensitive to the choice of the distance parameter. We chose to normalize the Euclidean distances or use (1 – *cosine*) in order to suppress unwanted effects of size on the distance, or effects of large numbers of zeros on the similarity. However, these two measures correlated negatively, and the results from the one based on relative Euclidean distances—the most natural distance measure—were difficult to interpret. The use of the construct (1 – *cosine*) as a distance measure is also debatable because, as Brandes (2008, at p. 142) noted, the topology of rescaled values might not reflect the intuition of distances between vertices.

One conceptual advantage of the Rao-Stirling diversity measure over betweenness centrality as used in this study is that the values are not binarized during the computation of diversity. An algorithm that would weigh the *cosine* values as a basis for the computation of betweenness centrality would perhaps improve our capacity to indicate interdisciplinarity (Brandes, 2001). Other variants of betweenness centrality such as flow-betweenness could further be explored (Brandes, 2008; Brandes & Fleischer, 2006; Newman, 2005).

In summary, one of the key findings of this study is that different indicators may capture different understandings of such a multi-faceted concept as interdisciplinarity. This is illustrated, for example, by the three factors found in Table 9. Similar results suggesting that more than a single indicator is needed to cover "interdisciplinarity" were found by Kajikawa *et al.* (2009) in a validation study, and theoretically specified as an expectation by Wagner *et al.* (2009; NSB, 2010, at p. 5-35). Such an elaboration of the conceptualization and operationalization of "interdisciplinarity" might be akin to the different indicators distinguished for measuring centrality in network analysis, with each of them capturing a particular aspect of this concept (Freeman, 1978/9).

**Acknowledgement**

We are grateful to Ulrik Brandes, Wouter de Nooy, and Andrew Stirling for suggestions and comments. Ismael Rafols acknowledges support from the US National Science Foundation (Award #0830207, "Measuring and Tracking Research Knowledge Integration").